\documentclass[pre,twocolumn]{revtex4-1}

\usepackage{graphicx}
\usepackage{color}
\usepackage{epstopdf}
\usepackage{amsmath,amssymb,bm,mathrsfs}

\newcommand{\Brho}{\boldsymbol\rho}

\renewcommand{\Im}{{\mathrm {Im}}}

\newcommand{\lap}{\Delta}

\newcommand{\grad}{\nabla}

\newcommand{\Br}{{\bf r}}

\newcommand{\Bn}{{\bf n}}
\newcommand{\Bu}{{\bf u}}
\newcommand{\Bp}{{\bf p}}
\newcommand{\Bx}{{\bf x}}
\newcommand{\By}{{\bf y}}

\newcommand{\Bq}{{\bf q}}
\newcommand{\Bk}{{\bf k}}
\newcommand{\Bv}{{\bf v}}

\newcommand{\BQ}{{\bf Q}}

\newcommand{\Bxi}{{\bm \xi}}

\begin{document}

\title{Acousto-optic effect in random media}

\author{Jeremy G. Hoskins}
\affiliation{Department of Mathematics, University of Michigan, Ann Arbor, MI 48109}
\email{jhoskin@umich.edu}

\author{John C. Schotland}
\affiliation{Department of Mathematics and Department of Physics, University of Michigan, Ann Arbor, MI 48109}
\email{schotland@umich.edu}

\date{\today}

\begin{abstract}
We consider the acousto-optic effect in a random medium. We derive the radiative transport equations that describe the propagation of multiply-scattered light in a medium whose dielectric permittivity is modulated by an acoustic wave. Using this result, we present an analysis of the sensitivity of an acousto-optic measurement to the presence of a small absorbing inhomogeneity. 
\end{abstract}

\maketitle

\section{Introduction}

The acousto-optic effect refers to the scattering of light from a medium whose optical properties are modulated by an acoustic wave. Brillouin scattering from density fluctuations in a fluid~\cite{Born-Wolf} and the ultrasonic modulation of multiply-scattered light~\cite{Leutz_1995} are familiar examples of this effect. It is well known that the scattered optical field carries information about the medium. This principle has been exploited to develop an imaging modality, known as acousto-optic imaging, which combines the spectroscopic sensitivity of optical methods with the spatial resolution of ultrasonic imaging. Two forms of acousto-optic imaging are usually distinguished. Direct imaging employs a focused ultrasound beam for image formation~\cite{Marks_1993,Kempe_1997,Granot_2001,
Wang_1995,Wang_1998_1,Wang_1998_2,Yao_2000,Li_2002_1,Li_2002_2,Leveque_1999,Leveque_2001,Atlan_2005,Gross_2009,Lev_2000,Lev_2002}. The image is created by scanning the focus of the beam and recording the intensity of the scattered light at a fixed detector. Tomographic imaging utilizes an inverse scattering method to reconstruct images of the optical properties of the medium~\cite{Bal_2010,varma_2011,Ammari_2014_1,Ammari_2014_2,Ammari_2013,BalSchotland_PhysRev2014,BalMoskow,BCS,chung}. 

The theory of the acousto-optic effect begins with a model for the propagation of electromagnetic waves in a material medium. The most general such model is based on the Maxwell equations for a dielectric whose permittivity is modulated by an acoustic wave~\cite{Born-Wolf}. Alternatively, for multiply-scattered light, a {phenomenological} theory based on the radiative transport equation (RTE) or the diffusion approximation (DA) to the RTE may be employed~\cite{mahan_1998,Bal_2010,hollman_2014,sakadzic_2006}. In this paper, we develop a \emph{first-principles} theory of the acousto-optic effect. We begin by constructing a model for the acoustic modulation of the dielectric permittivity of a medium consisting of small scatterers suspended in a fluid. Next, we consider the propagation of light in the medium and obtain the wave equations obeyed by the frequency components of the optical field at harmonics of the acoustic frequency. We then obtain the corresponding RTE by
asymptotic analysis of the Wigner transform of the field in a random medium. We note that the problem is challenging because the random medium acquires a time-dependence due to the presence of the acoustic field. We apply our results to estimating the minimum detectable size of a small inhomogeneity in acousto-optic imaging.

%Since the scatterers in the medium are displaced by the acoustic wave, the scattered light undergoes a frequency shift which permits the localization of the resulting so-called tagged photons to the volume containing the focus. 

The remainder of this paper is organized as follows. Our model for the acousto-optic effect is introduced in Sec.~II. In Sec.~III, we use this model to derive the RTE.
The corresponding DA is discussed in Sec.~IV. Sec.~V discusses the application of the obtained DA to the problem of detecting a small inhomogeneity. Our conclusions are formulated in Sec.~VI. Several appendices contain the technical details of long calculations.

\section{Acousto-optic effect}
In this section we develop a simple model for the acousto-optic effect. The setup we consider is illustrated in Figure~1.

\subsection{Model}
We begin by considering a medium consisting of identical neutrally-buoyant spherical particles suspended in a fluid. We suppose that an acoustic wave propagates in the fluid, the effect of which is to cause the particles to move under the associated radiation force. If the amplitude of the acoustic wave is sufficiently small, the particles will oscillate about their equilibrium positions. It is then possible to treat the motion of each particle as independent, neglecting hydrodynamic interactions. It follows that the equation of motion of a single particle is of the form
\begin{equation}
\label{eq_motion}
\varrho \frac{d \Bu}{dt} = - \grad p + \frac{4\pi a \eta}{V}(\Bv-\Bu) \ .
\end{equation} 
Here $\Bu$ denotes the velocity of the particle, $p$ is the pressure, $\Bv$ is the velocity field in the fluid, $\eta$ is the viscosity, $a$ is the radius of the particle, $\varrho$ is its mass density and $V=4\pi a^3/3$.  Consider a standing time-harmonic acoustic wave with pressure
\begin{equation}
p(\Bx,t) = p_0\cos(\Omega t) \cos(\BQ\cdot \Bx) \ ,
\end{equation}
where $p_0$ is the amplitude of the wave, $\Omega$ is its frequency and $\BQ$ is the wave vector. Here we have assumed that the speed of sound $c_s$ is constant with $Q=\Omega/c_s$.
The corresponding velocity field is given by
\begin{equation}
\Bv = \frac{p_0}{\varrho\Omega} \sin(\Omega t) \sin(\BQ\cdot\Bx) \BQ \ .
\end{equation}
Thus apart from a transient, the particle moves with the fluid.
Let $\Bx_1,\ldots,\Bx_N$ denote the positions of the particles and
\begin{equation} 
\rho(\Bx,t)=\sum_{j=1}^N\delta(\Bx-\Bx_j(t)) 
\end{equation}
their density. Since each particle is independent, it follows from integration of the equations of motion (\ref{eq_motion}) that $\rho$ is given by
\begin{equation}
\label{density}
\rho(\Bx,t) = \rho_0(\Bx)\left[1 + \delta \cos(\Omega t) \cos(\BQ\cdot\Bx)\right] \ ,
\end{equation}
where $\rho_0$ is the number density of the particles in the absence of the acoustic wave and $\delta = p_0/(\rho c_s^2)$ is a small parameter. Note that taking $\delta \ll 1$ is consistent with the neglect of hydrodynamic interactions. We conclude that the number density of particles is modulated by the acoustic wave.

\begin{figure}[t] 
\vspace{-.5in}
\begin{center}
\includegraphics[width=10cm]{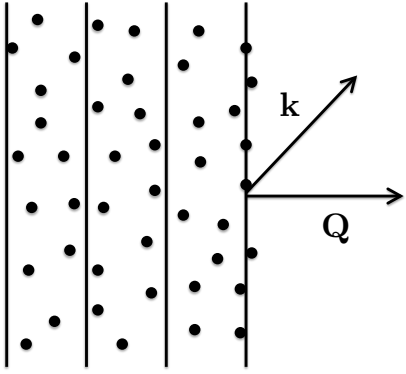}
\vspace{-0.6in} 
\end{center}
\caption{Illustrating the acousto-optic effect in a random medium.}
\end{figure}

Next, we turn to the propagation of light in the medium. For simplicity, we ignore the effects of polarization and employ a scalar theory of the optical field. The field $u(\Bx,t)$ is taken to obey the wave equation
\begin{equation}
\label{wave_eqn}
\frac{1}{c^2}\frac{\partial^2}{\partial t^2}\left(\varepsilon(\Bx,t) u \right) = \lap u \ ,
\end{equation}  
where $\varepsilon$ is the dielectric permittivity of the medium and $c$ is the speed of light in vacuum. The permittivity is of the form
\begin{equation}
\varepsilon(\Bx,t) = \varepsilon_0(\Bx,t) + 4\pi\eta(\Bx,t) \ ,
\end{equation}
where $\varepsilon_0$ is the permittivity of the fluid and $\eta$ is the dielectric susceptibility of the particles. The permittivity of the fluid is acoustically modulated and is given by
\begin{equation}
\medskip
\varepsilon_0(\Bx,t) = \varepsilon_0\left[1 + \delta \gamma \cos(\Omega t) \cos(\BQ\cdot\Bx)\right] \ ,
\end{equation}
where $\varepsilon_0$ is the permittivity of the fluid in the absence of the acoustic wave and $\gamma$ is the elasto-optical constant~\cite{Born-Wolf}. The fluid is taken to be nonabsorbing, so that $\varepsilon_0$ is purely real, positive and frequency independent. In addition, we suppose that the particles are small in size in comparison to the wavelength of light. That is, we treat the particles as point scatterers~\cite{deVries_1998}. The susceptibility is then given by $\eta(\Bx,t) = \alpha\rho(\Bx,t)$, where $\alpha$ is the polarizability of a single particle. Using (\ref{density}), we see that $\eta(\Bx,t)$ is given by
\begin{equation}
\eta(\Bx,t) = \eta(\Bx)\left[1 + \delta \cos(\Omega t) \cos(\BQ\cdot\Bx)\right] \ ,
\end{equation}
where $\eta(\Bx)=\alpha\rho_0(\Bx)$.

We suppose that the field $u$ is monochromatic with frequency $\omega$ and time-dependence $e^{-i\omega t}$. It will prove useful to decompose $u$ in harmonics of the acoustic frequency according to
\begin{equation}
\label{coupled}
u(\Bx,t) = \sum_{n=-\infty}^\infty e^{-i(\omega + n\Omega)t} u_n(\Bx) \ .
\end{equation}
It follows from (\ref{wave_eqn}) that the Fourier components $u_n$ obey the system of coupled Helmholtz equations
\begin{align}
\nonumber
\lap u_n +  k_n^2 \left(\varepsilon_0 + 4\pi \eta(\Bx)\right) u_n = \\
-\frac{\delta k_n^2}{2}
\left(\gamma\varepsilon_0 + 4\pi \eta(\Bx)\right)\cos(\BQ\cdot\Bx)\left(u_{n-1} + u_{n+1}\right) \ , 
\end{align}
where $k_n = (\omega + n \Omega)/c$. Note that if $u_0 = O(1)$ then $u_n = O(\delta^n)$. 
Here we do not consider modes $u_n$ for $|n| \ge 2$ and close the equations (\ref{coupled})
as
\begin{widetext}
\begin{eqnarray}
\label{u0}
\lap u_0 + k_0^2 \left(\varepsilon_0 + 4\pi \eta(\Bx)\right) u_0 &=& -\frac{\delta k_0^2}{2}
\left(\gamma\varepsilon_0 + 4\pi \eta(\Bx)\right)\cos(\BQ\cdot\Bx)\left(u_{-1} + u_{1}\right) \ , \\
\lap u_{1} +  k_1^2 \left(\varepsilon_0 + 4\pi \eta(\Bx)\right) u_1 &=& -\frac{\delta k_1^2}{2}
\left(\gamma\varepsilon_0 + 4\pi \eta(\Bx)\right)\cos(\BQ\cdot\Bx)u_{0} \ , \\
\lap u_{-1} +  k_{-1}^2 \left(\varepsilon_0 + 4\pi \eta(\Bx)\right) u_{-1} &=& -\frac{\delta k_{-1}^2}{2}\left(\gamma\varepsilon_0 + 4\pi \eta(\Bx)\right)\cos(\BQ\cdot\Bx)u_{0} \ .
\end{eqnarray}
Furthermore, the right hand side of (\ref{u0}) can be neglected since it is $O(\delta^2)$. 
The above equations thus become
\begin{eqnarray}
\label{u0_final}
\lap u_0 + k_0^2 \left(\varepsilon_0 + 4\pi \eta(\Bx)\right) u_0 &=& 0 \ , \\
\label{u1}
\lap u_{1} +  k_1^2 \left(\varepsilon_0 + 4\pi \eta(\Bx)\right) u_1 &=& -\frac{\delta k_1^2}{2}
\left(\gamma\varepsilon_0 + 4\pi \eta(\Bx)\right)\cos(\BQ\cdot\Bx)u_{0} \ , \\
\lap u_{-1} +  k_{-1}^2 \left(\varepsilon_0 + 4\pi \eta(\Bx)\right) u_{-1} &=& -\frac{\delta k_{-1}^2}{2}\left(\gamma\varepsilon_0 + 4\pi \eta(\Bx)\right)\cos(\BQ\cdot\Bx)u_{0} \ .
\label{u-1}
\end{eqnarray}
\end{widetext}
Note that in this form, the equations for $u_0$ and $u_{\pm 1}$ are decoupled.  For the remainder of the paper, we will take (\ref{u0_final})--(\ref{u-1}) to be the equations governing the acousto-optic effect.

\begin{table}[b]
\begin{center}
\begin{tabular}{ l | c | c }
\hline
Name & Symbol & Value \\ \hline
\hline Propagation distance & $L$ & 1 cm \\ 
\hline Acoustic frequency & $\Omega$ & $10^6$ Hz \\ 
\hline Optical frequency & $\omega$ & $10^{15}$ Hz \\
\hline Speed of sound & $c_s$ & $1.5 \times 10^5$ cm s$^{-1}$ \\ 
\hline Mass density & $\varrho$ & 1 g cm$^{-3}$ \\ 
\hline Pressure amplitude & $p_0$ &  $10^6$ g cm$^{-1}$ s$^{-2}$ \\ 
\hline Absorption coefficient & $\mu_a$ &  $0.1$ cm$^{-1}$  \\ 
\hline Scattering coefficient & $\mu_s$ &  $10$ cm$^{-1}$  \\ 
\hline Transport mean free path & $\ell^*$ &  $0.1$ cm  \\ \hline
\hline \hspace{40pt} $\delta$ & $p_0/(\varrho c_s^2)$ & $O(10^{-4})$  \\ 
\hline \hspace{40pt} $\epsilon$ & $\lambda/L$ &  $O(10^{-4})$ \\ 
\hline 
\end{tabular} =-`1   
\caption{Values of parameters arising in the acousto-optic effect. The numbers chosen are representative of biological tissue.}
\end{center}
\end{table}

\subsection{Homogeneous medium}
We now consider the case of a homogeneous fluid medium and put the susceptibility of the particles $\eta \equiv 0$. Evidently, the fundamental mode $u_0$ acts as a source of the first harmonics $u_{\pm 1}$. In addition, the modes $u_{\pm 1}$ are independent. Note that $u_{-1}$ can be obtained from $u_1$ by performing
the replacement $k_1 \to k_{-1}$. The solution to (\ref{u1}) is given by
\begin{equation}
u_1(\Bx) =  \frac{1}{2} \gamma \varepsilon_0 \delta k_1^2  \int d^3x' G(\Bx,\Bx')\cos(\BQ\cdot\Bx') u_0(\Bx') \ .
\end{equation}
Here the Green's function $G$, which obeys the equation
\begin{equation}
\lap_\Bx G(\Bx,\Bx') + \varepsilon_0 k_1^2 G(\Bx,\Bx') = -\delta(\Bx-\Bx') \ ,
\end{equation}
is given by
\begin{equation}
G(\Bx,\Bx') = \frac{e^{i\sqrt{\varepsilon_0}k_1|\Bx-\Bx'|}}{4\pi|\Bx-\Bx'|} \ .
\end{equation}
If the field $u_0$ is a unit-amplitude plane wave of the form
\begin{equation}
u_0(\Bx) = e^{i\Bk \cdot \Bx} \ , \quad k=\sqrt{\varepsilon_0}k_0 \ ,
\end{equation}
we find that $u_1$ is given by
\begin{eqnarray}
\label{u_1}
\nonumber
u_1(\Bx) = \frac{1}{4}\delta k'^2 \Bigg[\frac{1}{(\BQ+\Bk)^2-k'^2}e^{i(\Bk+\BQ)\cdot\Bx} \\
 + \frac{1}{(\BQ-\Bk)^2-k'^2} e^{i(\Bk-\BQ)\cdot\Bx}\Bigg] \ ,
\end{eqnarray}
where $k'=\sqrt{\varepsilon_0}k_1$. Evidently for fixed $\BQ$, there is a resonance if  the incident wave vector $\Bk$ obeys the condition
\begin{equation}
\label{resonance}
|\BQ \pm \Bk| = k' \ .
\end{equation}
Since $\Omega \ll \omega$, it follows that $k\simeq k'$. Thus (\ref{resonance}) becomes
\begin{equation}
\left(\frac{\BQ}{2}\right)^2 = \pm \Bk\cdot \frac{\BQ}{2} \ ,
\end{equation}
which we recognize as the Bragg condition~\cite{Born-Wolf}. 
Equivalently,
\begin{equation}
\label{resonance_angle}
\cos\theta = \pm\frac{Q}{2k} \ ,
\end{equation}
where $\theta$ is the angle between $\Bk$ and $\BQ$. That is, a resonance occurs for $\theta\simeq \pm\pi/2$. We note that the presence of absorption in the fluid prohibits the formation of a resonance. That is, the denominators in (\ref{u_1}) can never vanish if $\varepsilon_0$ acquires even a small imaginary part.

%We note that neglecting the right hand side of (\ref{u0}) is not justified near resonance. 

% analysis of resonances in random media.

\bigskip

\section{Radiative transport}
We now turn to the theory of the acousto-optic effect in random media. We take the modes $u_0$ and $u_1$ to obey (\ref{u0_final}) and (\ref{u1}), and assume that the susceptibility $\eta$ is a random field with correlations
\begin{eqnarray}
\langle \eta \rangle &=& 0 \ , \\
\langle \eta(\Bx) \eta(\Bx') \rangle &=& C(\Bx-\Bx') \ ,
\label{def_C}
\end{eqnarray}
where $\langle \cdots \rangle$ denotes statistical averaging. We assume that the medium is statistically homogeneous and isotropic. That is, the correlation function $C(\Bx-\Bx')$ depends only upon the quantity $|\Bx-\Bx'|$. 

To make further progress, we must consider the relative sizes of the important physical scales. This leads us to introduce two small parameters: $\delta=p_0/(\varrho c_s^2)$ and $\epsilon = 1/(k_0L)$, where $L$ is the distance over which the optical field propagates. According to Table~I, we see that $\epsilon=O(\delta)$. Henceforth, we will put $\delta=\epsilon$, which can always be arranged by adjusting the strength of the amplitude $p_0$. Now, the solutions of (\ref{u0_final}) and (\ref{u1}) oscillate on the scale of the optical wavelength $\lambda=2\pi/k_0$. However, we are interested in the behavior of the solutions on the macroscopic scale $L\gg \lambda$. To this end, we rescale the position $\Bx$ by $\Bx \to \Bx/\epsilon$ with $\epsilon \ll 1$. In addition, we assume that the randomness is sufficiently
weak so that the correlation function $C$ is of the order $O(\epsilon)$. Thus (\ref{u0_final}) and (\ref{u1}) become
%\begin{widetext}
\begin{eqnarray}
\label{u0_rescaled}
\epsilon^2\lap u_\epsilon + k_0^2 \left(\varepsilon_0 + 4\pi \sqrt\epsilon\eta(\Bx/\epsilon)\right) u_\epsilon &=& 0 \ ,\\
\nonumber
\label{u1_rescaled}
\epsilon^2 \lap v_{\epsilon} +  k_1^2 \left(\varepsilon_0 + 4\pi \sqrt\epsilon\eta(\Bx/\epsilon)\right) v_\epsilon &=& \\ \hspace{-40pt} -\frac{\epsilon k_1^2}{2} 
\left(\gamma\varepsilon_0 + 4\pi \sqrt\epsilon\eta(\Bx/\epsilon)\right)
\cos(\BQ\cdot\Bx)u_\epsilon
\ , 
\end{eqnarray}
%\end{widetext}
where $u_\epsilon(\Bx) = u_0(\Bx/\epsilon)$ and $v_\epsilon(\Bx) = u_1(\Bx/\epsilon)$.
Note that we have rescaled the susceptibility $\eta$ by $\eta \to \sqrt\epsilon\eta$ to be consistent with the condition that $C$ is $O(\epsilon)$. We also note that we have not rescaled the term $\cos(\BQ\cdot\Bx)$ since it is slowly varying on the scale of the optical wavelength. That is, the random medium does not vary on the same scale as the periodic modulation of the fluid. It will prove useful to rewrite (\ref{u0_rescaled}) and (\ref{u1_rescaled}) in the form
%\begin{widetext}
\begin{eqnarray}
\label{phi_helmholtz}
\nonumber
\epsilon^2\lap \phi_\epsilon + k_0^2\left(\varepsilon_0 + 4\pi \sqrt\epsilon\eta(\Bx/\epsilon)\right) \phi_\epsilon = \\
-\frac{\epsilon k_0^2}{2}
\left(\gamma\varepsilon_0 + 4\pi \sqrt\epsilon\eta(\Bx/\epsilon)\right)\cos(\BQ\cdot\Bx)A\phi_\epsilon \ ,
\end{eqnarray}
%\end{widetext}
where
\begin{equation}
A = \left(\begin{array}{cc}0 & 0 \\  1 & 0\end{array}\right) \ , \quad
\phi_\epsilon = \left(\begin{array}{c} u_\epsilon \\ v_\epsilon \end{array}\right) \ ,
\end{equation}
and since $\Omega \ll \omega$, we have made the approximation $k_0\simeq k_1$.

We now turn to the derivation of the RTE.
The Wigner transform of $\phi_\epsilon$ is defined as 
\begin{equation}
W_\epsilon(\Bx,\Bk) = \int \frac{d^3 x'}{(2\pi)^3} e^{i\Bk\cdot\Bx'} \phi_\epsilon(\Bx-\epsilon\Bx'/2)\phi_\epsilon^\dag(\Bx+\epsilon\Bx'/2) \ .
\end{equation}
The Wigner transform is a $2\times2$ Hermitian matrix that is related to the energy density and energy current of the modes by 
\begin{align}
\phi_\epsilon(\Bx) \phi_\epsilon^\dag(\Bx) &= \int W_\epsilon(\Bx,\Bk) d^3k \ , \quad\quad \\
\nonumber
\frac{i\epsilon}{2}\left[ \phi_\epsilon(\Bx) \grad \phi_\epsilon^\dag(\Bx) -\grad\phi_\epsilon(\Bx) \phi_\epsilon^\dag(\Bx)\right] &= \int \Bk W_\epsilon(\Bx,\Bk) d^3k \ . \\
\end{align}
We will see that the Wigner transform plays the role of a phase-space energy density. We note that the Wigner transform is not directly measurable. Nevertheless, in the $\epsilon\to 0$ limit, the average of $W_\epsilon$ may be interpreted as the specific intensity in radiative transport theory. 

It can be shown that $W_\epsilon$ obeys the Liouville equation
\begin{eqnarray}
\label{liouville}
\Bk\cdot \grad_\Bx W_\epsilon + \left(\frac{1}{\sqrt\epsilon} L_1 + L_2 +\sqrt\epsilon L_3 \right) W_\epsilon = 0 \ .
\end{eqnarray}
Here the operators $L_1$, $L_2$ and $L_3$ are defined by
\begin{widetext}
\begin{align}
\label{L1}
&L_1W_\epsilon(\Bx,\Bk) =  2\pi i k_0^2\int \frac{d^3p}{(2\pi)^3} e^{i{\bf p}\cdot{\bf x}/\epsilon}\hat{\eta}({\bf p}) \left[W_\epsilon\left({\bf x},{\bf k}+\frac{\bf p}{2}\right)-W_\epsilon\left({\bf x},{\bf k}-\frac{\bf p}{2}\right)\right] \ , \\ 
\label{L2}
&L_2W_\epsilon(\Bx,\Bk) = \frac{i}{8}\gamma\varepsilon_0 k_0^2 A\left[ e^{i{\bf Q}\cdot {\bf x}}W_\epsilon\left({\bf x},{\bf k}+\frac{\epsilon}{2}{\bf Q}\right) + e^{-i{\bf Q}\cdot {\bf x}}W_\epsilon\left({\bf x},{\bf k}-\frac{\epsilon}{2}{\bf Q}\right) \right] \\
\nonumber
& \hspace{2cm} -\frac{i}{8}\gamma\varepsilon_0 k_0^2\left[ e^{-i{\bf Q}\cdot {\bf x}}W_\epsilon\left({\bf x},{\bf k}+\frac{\epsilon}{2}{\bf Q}\right) + e^{i{\bf Q}\cdot {\bf x}}W_\epsilon\left({\bf x},{\bf k}-\frac{\epsilon}{2}{\bf Q}\right) \right]A^\dag \ , \\
\label{L3}
&L_3W_\epsilon(\Bx,\Bk) =  \frac{i}{2}\pi k_0^2  \int\frac{d^3p}{(2\pi)^3} e^{i{\bf p}\cdot{\bf x}/\epsilon} \hat{\eta}({\bf p})A\left[e^{i{\bf Q}\cdot {\bf x}}W_\epsilon\left({\bf x},{\bf k}+\frac{\bf p}{2}+\frac{\epsilon}{2}{\bf Q}\right)+e^{-i{\bf Q}\cdot {\bf x}}W_\epsilon\left({\bf x},{\bf k}+\frac{\bf p}{2}-\frac{\epsilon}{2}{\bf Q}\right) \right] \\
\nonumber
&-\frac{i}{2}\pi k_0^2 \int \frac{d^3p}{(2\pi)^3} e^{i{\bf p}\cdot{\bf x}/\epsilon} \hat{\eta}({\bf p})\left[e^{-i{\bf Q}\cdot {\bf x}}W_\epsilon\left({\bf x},{\bf k}+\frac{\bf p}{2}+\frac{\epsilon}{2}{\bf Q}\right)+e^{i{\bf Q}\cdot {\bf x}}W_\epsilon\left({\bf x},{\bf k}+\frac{\bf p}{2}-\frac{\epsilon}{2}{\bf Q}\right) \right] A^\dag \ .
\end{align}
\end{widetext}
See Appendix A for the derivation of the above result. 

We now consider the asymptotics of the Wigner transform in the high-frequency limit $\epsilon\to 0$. We will see that averaging over realizations of the random medium leads directly to the required radiative transport equations. Following standard procedures~\cite{Ryzhik_1996,Caze_2015}, we introduce a multiscale expansion for the Wigner transform of the form
\begin{equation}
\label{multiscale}
W_\epsilon(\Bx,\Bk) = W_0(\Bx,\Bk) + \sqrt\epsilon W_1(\Bx,\Bxi,\Bk) + \epsilon W_2(\Bx,\Bxi,\Bk) + \cdots \ ,
\end{equation}
where $\Bxi=\Bx/\epsilon$ is a fast variable and $W_0$ is taken to be deterministic. We then regard $\Bx$ and  $\Bxi$ as independent and make the replacement
\begin{equation}
\grad_\Bx \to \grad_\Bx + \frac{1}{\epsilon}\grad_\Bxi \ .
\end{equation}
Eq.~(\ref{liouville}) thus becomes
\begin{align}
\label{liouville_again}
\epsilon \Bk\cdot\grad_\Bx W_\epsilon + \Bk\cdot\grad_\Bxi W_\epsilon
+ \left(\sqrt\epsilon L_1 + \epsilon L_2 + \epsilon^{3/2}L_3\right)W_\epsilon = 0 \ .
\end{align}
Substituting (\ref{multiscale}) into (\ref{liouville_again}) and collecting terms of
order $O(\sqrt\epsilon)$, we obtain
\begin{equation}
\Bk\cdot\grad_\Bxi W_1 + L_1W_0 = 0 \ .
\end{equation}
The above equation is readily solved for $W_1$ with the result
\begin{equation}
\label{W1}
\widetilde W_1(\Bx,\Bq,\Bk) = 2\pi k_0^2\tilde\eta(\Bq) \frac{W_0(\Bx,\Bk+\Bq/2)-W_0(\Bx,\Bk-\Bq/2)}{\Bq\cdot\Bk + i\theta} \ .
\end{equation}
Here the Fourier transform of $W_1$ is defined by
\begin{equation}
\widetilde W_1(\Bx,\Bq,\Bk) = \int d^3\xi e^{-i\Bq\cdot\Bxi} W_1(\Bx,\Bxi,\Bk)
\end{equation}
and $\theta$ is a small positive regularization parameter that will be set to zero later in the calculation. 

At $O(\epsilon)$ we find that
\begin{equation}
\label{O(epsilon)}
\Bk\cdot\grad_\Bx W_0 + \Bk\cdot\grad_\Bxi W_2 + L_1W_1 + L_2 W_0  = 0 \ ,
\end{equation}
where $L_2W_0$, as defined in (\ref{L2}), is evaluated at $\epsilon=0$.
The RTE may be derived by averaging (\ref{O(epsilon)}) over the random field $\eta$. To proceed, we make the assumption $\langle\Bk\cdot\grad_\Bxi W_2\rangle=0$, 
which is consistent with the stationarity of $W_2(\Bx,\Bxi,\Bk)$ in $\Bxi$. We find that (\ref{L2}) becomes
\begin{widetext}
\begin{align}
\label{before_averaging}
\Bk\cdot\grad_\Bx W_0 + 2\pi i k_0^2 \int \frac{d^3p}{(2\pi)^3} e^{i\Bp\cdot\Bx/\epsilon}
\langle\tilde\eta(\Bp)\left[W_1(\Bx,\Bq,\Bk+\Bp/2) - W_1(\Bx,\Bq,\Bk-\Bp/2)\right]\rangle
+\frac{i}{4}\gamma\varepsilon_0 k_0^2\cos(\BQ\cdot\Bx)\left(AW_0 - W_0A^\dag \right) = 0 \ ,
\end{align}
\end{widetext}
where we have used the fact that $W_0$ is deterministic. Next, we substitute the expression (\ref{W1}) for $W_1$ into (\ref{before_averaging}) and, upon carrying out the indicated average,
we obtain
\begin{align}
\label{pre_RTE}
\nonumber
&\Bk\cdot\grad_\Bx W_0 +\frac{i}{4}\gamma\varepsilon_0 k_0^2\cos(\BQ\cdot\Bx)\left(AW_0 - W_0A^\dag \right) \\
&=k_0^4\int d^3p \delta\left(\frac{p^2}{2}- \frac{k^2}{2}\right)\tilde C(\Bp-\Bk)\left[W_0(\Bx,\Bp) - W_0(\Bx,\Bk)\right] \ .
\end{align} 
%\end{widetext}
See Appendix B for the details of this calculation. Note that the presence of the delta function in the above result indicates that $W_0$ depends only upon the direction $\hat\Bk$. It is then convenient to define the specific intensity $I$, phase function $f$ and scattering coefficient $\mu_s$ by
\begin{align}
&\delta(k-k_0)I(\Bx,\hat\Bk) = W_0(\Bx,k\hat\Bk) \ , \\
& f(\hat\Bk,\hat\Bk') = 
\frac{\tilde C(k_0(\hat\Bk-\hat\Bk'))}{\displaystyle\int \tilde C(k_0(\hat\Bk-\hat\Bk'))d\hat\Bk'} \ , \\
&\mu_s = k_0^4 \int \tilde C(k_0(\hat\Bk-\hat\Bk'))d\hat\Bk' \ .
\end{align}
Making use of the above definitions, we find that (\ref{pre_RTE}) becomes 
\begin{align}
\label{RTE}
\nonumber
&\hat\Bk\cdot\grad_\Bx I + \mu_s I + \frac{i}{4}\gamma\varepsilon_0 k_0\cos(\BQ\cdot\Bx)\left(AI - 
IA^\dag \right) \\
&=\mu_s\int d\hat\Bk' f(\hat\Bk,\hat\Bk')I(\Bx,\hat\Bk') \ .
\end{align} 
We note that $\mu_s$ and $f$ are given in terms of correlations of the medium. Since the susceptibility $\eta$ is statistically homogeneous and isotropic, $\tilde C(k_0(\hat\Bk-\hat\Bk'))$ depends only on the quantity $|\hat\Bk-\hat\Bk'|$, and therefore $f(\hat\Bk,\hat\Bk')$ depends solely on $\hat\Bk\cdot\hat\Bk'$. Likewise, $\mu_s$ does not depend on the direction $\hat\Bk$. Finally, we point out that in the case of white noise disorder, the correlation function $C(\Bx) = C_0\delta(\Bx)$, where $C_0$ is constant. We find that
\begin{equation}
\mu_s = 4\pi k_0^4 C_0 \ , \quad f = 1/(4\pi) \ ,
\end{equation}
which corresponds to isotropic scattering. More generally, if the medium consists of identical discrete scatterers,  $\mu_s$ and $A$ are related to the total scattering cross section and differential scattering cross section, respectively~\cite{Caze_2015}.

Eq.~(\ref{RTE}) can be expressed as a system of coupled equations of the form
\begin{align}
\label{RTE_0}
\hat\Bk\cdot \grad_\Bx I_{00} + \mu_s I_{00} - \mu_s LI_{00} & = 0 \ , \\
\hat\Bk\cdot \grad_\Bx I_{01} + \mu_s I_{01} - \mu_s LI_{01} & = \frac{i}{4}\gamma\varepsilon_0 k_0\cos(\BQ\cdot\Bx)I_{00} \ , \\
\hat\Bk\cdot \grad_\Bx I_{10} + \mu_s I_{10} - \mu_s LI_{10} & = -\frac{i}{4}\gamma\varepsilon_0 k_0\cos(\BQ\cdot\Bx)I_{00} \ , \\
\hat\Bk\cdot \grad_\Bx I_{11} + \mu_s I_{11} - \mu_s LI_{11} & = \frac{1}{2}\gamma\varepsilon_0 k_0 \cos(\BQ\cdot\Bx)\Im  I_{01} \ ,
\label{RTE_1}
\end{align}
where the operator $L$ is defined by
\begin{equation}
LI(\Bx,\hat\Bk) = \int f(\hat\Bk,\hat\Bk')I(\Bx,\hat\Bk')d\hat\Bk' \ .
\end{equation}
Eqs.~(\ref{RTE_0})--(\ref{RTE_1}) are the main result of this paper. They may be understood as a system of RTEs that describe the acousto-optic effect in random media. The quantity $I_{00}$ is the specific intensity of light at the fundamental frequency and (\ref{RTE_0}) is the corresponding RTE. Similarly, $I_{11}$ is the specific intensity of the first harmonic; it obeys the RTE (\ref{RTE_1}). We note that $I_{01}$ and $I_{10}$ are related to correlations of the modes $u_0$ and $u_1$.

\section{Diffusion Approximation}

In this section, we consider the diffusion limit of the radiative transport theory developed in Section~III. We begin by recalling the diffusion approximation (DA) for a RTE of the form
\begin{equation}
\hat\Bk\cdot\grad_\Bx I + (\mu_a + \mu_s)I -\mu_sLI = Q \ ,
\end{equation}
where $\mu_a$ is the absorption coefficient and $Q$ is the source. The DA is obtained by expanding $I$ in angular harmonics~\cite{Duderstadt-Martin}. To lowest order, it can be seen that
\begin{equation}
I(\Bx,\hat\Bk) = \frac{1}{4\pi}\left(U(\Bx) -\ell^* \hat\Bk\cdot\grad U(\Bx)\right) \ .
\end{equation}
Here the energy density $U$ obeys the diffusion equation
\begin{equation}
\label{DA}
-\frac{1}{3}\ell^* \lap U + \mu_a U = S \ ,
\end{equation}
Here the source $S=\int Q d\hat\Bk$ and the transport mean free path $\ell^*$ is defined by
\begin{equation}
\ell^* = 1/\left[(1-g)\mu_s + \mu_a\right] \ , \quad g = \int \hat\Bk\cdot\hat\Bk' f(\hat\Bk,\hat\Bk')d\hat\Bk' \ ,
\end{equation}
where $g$ is the anisotropy of scattering. We note that $-1\le g \le 1$ and $g=0$ for isotropic scattering. The DA holds when $\ell^*|\grad U| \ll U$ and breaks down in optically thin layers, in weakly scattering or strongly absorbing media, and near boundaries.

\begin{figure}[t] 
\vspace{-1.2in}
\begin{center}
\includegraphics[width=0.5\textwidth]{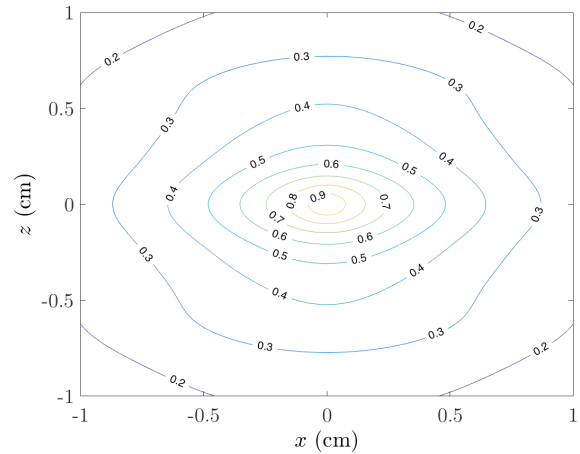}
\vspace{-1.2in}
\end{center}
\caption{Contour plot of $U_{11}$ in the $y=0$ plane.}
\end{figure}

Using the above results, we can immediately construct the DA for (\ref{RTE_0})--(\ref{RTE_1}). We find that
\begin{align}
I_{00}(\Bx,\hat\Bk) &= \frac{1}{4\pi}\left(U_{00}(\Bx) -\ell^* \hat\Bk\cdot\grad U_{00}(\Bx)\right) \ , \\
I_{01}(\Bx,\hat\Bk) &= \frac{1}{4\pi}\left(U_{01}(\Bx) -\ell^* \hat\Bk\cdot\grad U_{01}(\Bx)\right) \ , \\
%I_{10}(\Bx,\hat\Bk) = \frac{1}{4\pi}\left(U_{10}(\Bx) -\ell^* \hat\Bk\cdot\grad U_{10}(\Bx)\right) \ , \\
I_{11}(\Bx,\hat\Bk) &= \frac{1}{4\pi}\left(U_{11}(\Bx) -\ell^* \hat\Bk\cdot\grad U_{11}(\Bx)\right) \ .
\end{align}
Here the corresponding energy densities obey
\begin{align}
\label{DA0}
-\frac{1}{3}\ell^* \lap U_{00} + \mu_a U_{00} &= \delta(\Bx-\Bx_0) \ , \\
-\frac{1}{3}\ell^* \lap U_{01} + \mu_a U_{01} &= \frac{i}{4}\gamma\varepsilon_0 k_0\cos(\BQ\cdot\Bx)U_{00} \ , \\
%-\frac{1}{3}\ell^* \lap U_{10} + \mu_a U_{10} = -\frac{i}{4}\gamma\varepsilon_0 k_0\cos(\BQ\cdot\Bx)U_{00} \ , \\
-\frac{1}{3}\ell^* \lap U_{11} + \mu_a U_{11} &= \frac{1}{2}\gamma\varepsilon_0 k_0\cos(\BQ\cdot\Bx)\Im U_{01}\ ,
\label{DA1}
\end{align}
where $\Bx_0$ is the position of a unit-amplitude point source. 
Since $U_{10}=U_{01}^*$, we have omitted the diffusion equation obeyed by $U_{10}$ and the corresponding specific intensity $I_{10}$. Note that we have introduced the absorption coefficient $\mu_a$ in (\ref{DA0})--(\ref{DA1}) by hand.  This is necessary to regularize the divergence arising from the scattering resonance.

In an infinite homogeneous medium, the energy density $U_{00}$ due to a unit amplitude point source located at $\Bx_0$ is given by
\begin{equation} 
U_{00}(\Bx) = \frac{3}{4\pi \ell^*} \frac{e^{-\kappa |\Bx-\Bx_0|}}{|\Bx-\Bx_0|}  \ ,
\end{equation}
where $\kappa = \sqrt{3\mu_a/\ell^*}$. Using this result, we find that $U_{01}$ is given by
\begin{equation}
\label{int_U01}
U_{01}(\Bx) = \frac{i}{4}\gamma\varepsilon_0 k_0\int d^3x' G_D(\Bx,\Bx')\cos(\BQ\cdot\Bx')U_{00}(\Bx') \ ,
\end{equation}
where
%Here the diffusion Green's function $G_D$ is of the form
% which obeys
%\begin{equation}
%-\frac{1}{3}\ell^* \lap_\Bx G_D(\Bx,\Bx') + \mu_a G_D(\Bx,\Bx') = \delta(\Bx-\Bx') \ ,
%\end{equation}
\begin{equation}
G_D(\Bx,\Bx') = \frac{3}{4\pi \ell^*} \frac{e^{-\kappa |\Bx-\Bx'|}}{|\Bx-\Bx'|} 
\end{equation}
is the diffusion Green's function.
Carrying out the above integration, we obtain
\begin{equation}
\label{U01}
U_{01}(\Bx) = i\frac{\gamma\varepsilon_0 k_0}{4\pi \kappa \ell^{*2}} e^{-\kappa |\Bx-\Bx_0|}\cos(\BQ\cdot\Bx)\left[1 + 
O\left(Q|\Bx-\Bx_0|\right)\right] \ .
\end{equation}
The above formula holds in the regime of low acoustic frequency. It follows from (\ref{DA1}) that $U_{11}$ is given by
\begin{equation}
\label{int_U11}
U_{11}(\Bx) = \frac{1}{2}\gamma\varepsilon_0 k_0\int d^3x' G_D(\Bx,\Bx')\cos(\BQ\cdot\Bx')\Im U_{01}(\Bx') \ .
\end{equation}
Using (\ref{U01}) and performing the indicated integration, we find that
\begin{equation}
U_{11}(\Bx) =\frac{(\gamma\varepsilon_0 k_0)^2}{4\pi(\kappa \ell^*)^3} e^{-\kappa|\Bx-\Bx_0|}
\cos^2(\BQ\cdot\Bx)\left[1+O(Q|\Bx-\Bx_0|)\right] \ .
\end{equation}
Note that when $\kappa\to 0$, which corresponds to a nonabsorbing medium, the above formulas for $U_{01}$ and $U_{11}$ exhibit a divergence. 

A more careful evaluation of the integral (\ref{int_U01}) yields
\begin{equation}
U_{01}(\Bx) =  \frac{i\gamma\varepsilon_0k_0}{64\pi}\sum_{m} e^{im\BQ\cdot\Bx_0}\int_0^1 {d}s\, e^{ism\BQ\cdot \Br} \frac{e^{-\beta(s)|\Bx-\Bx_0|}}{\beta(s)},
\end{equation}
where the sum is over $m\in\{-1,1\}$ and 
\begin{equation}
\beta(s) = \sqrt{\kappa^2+m^2 \BQ^2 s(1-s^2)}.
\end{equation} 
Using this result and (\ref{int_U11}) we obtain
\begin{align}
\label{U11_integral}
\nonumber
& U_{11}(\Bx) = \frac{(\gamma \varepsilon_0 k_0)^2}{512 \pi}\sum_{m,n} e^{i\BQ\cdot(m\Bx+n\Bx_0)} \int_0^1 {d}s\, \int_0^{1-s}{d}t  \\
& \times e^{-i\Bp(s,t) \cdot \Br}	\frac{e^{-\phi(s,t) |\Bx-\Bx_0|}}{\phi^2(s,t)}\left(|\Bx-\Bx_0|+\frac{2}{\phi(s,t)}\right),
\end{align}
where 
\begin{align}
& \Bp(s,t) = (ms-nt)\BQ \ , \\
& \phi(s,t) =\sqrt{ \kappa^2+(s+t)\BQ^2-p^2(s,t) } \ .
\end{align}
See Appendix C for the derivation of (\ref{U11_integral}).
In Figure~2 we show a contour plot of the energy density $U_{11}$  for a source located at the origin. The physical parameters were chosen to be $Q = 10 \ {\rm cm}^{-1}$ and $\kappa = 1 \  {\rm cm}^{-1}$, which are typical in biomedical applications.

\section{Small absorbers}

In this section we consider the acousto-optic effect generated by a small absorbing inhomogeneity. As an application, we calculate the sensitivity of detection of the absorber. For simplicity, we work in the half-space geometry in which the optical source and detector are located on a planar boundary.

\subsection{Half-space geometry}
We consider a homogeneous medium that occupies the half-space $z<0$. The half-space $z>0$ is taken to be vacuum. In this setting, the energy densities $U_{00}$, $U_{01}$ and $U_{11}$ obey 
\begin{align}
%\label{DA0}
-\frac{1}{3}\ell^* \lap U_{00} + \mu_a U_{00} &= 0 \ , \\
-\frac{1}{3}\ell^* \lap U_{01} + \mu_a U_{01} &= \frac{i}{4}\gamma\varepsilon_0 k_0\cos(\BQ\cdot\Bx)U_{00} \ , \\
%-\frac{1}{3}\ell^* \lap U_{10} + \mu_a U_{10} = -\frac{i}{4}\gamma\varepsilon_0 k_0\cos(\BQ\cdot\Bx)U_{00} \ , \\
-\frac{1}{3}\ell^* \lap U_{11} + \mu_a U_{11} &= \frac{1}{2}\gamma\varepsilon_0 k_0\cos(\BQ\cdot\Bx)\Im U_{01}\ ,
%\label{DA1}
\end{align}
and satisfy the boundary conditions
\begin{align}
\label{boundary_source}
&U_{00}(\Bx) + \ell \hat{\Bn} \cdot \nabla U_{00}(\Bx) = S_0\delta(\Bx-\Bx_0) \ , \\
&U_{01}(\Bx) + \ell \hat{\Bn} \cdot \nabla U_{01}(\Bx) = 0 \ , \\
&U_{11}(\Bx) + \ell \hat{\Bn} \cdot \nabla U_{11}(\Bx) = 0 \ , 
\end{align}
on the plane $z=0$ with outward unit normal $\hat{\Bn}$. Here the parameter $\ell$ is the extrapolation distance and the right hand side of (\ref{boundary_source}) corresponds to a point source located on the boundary at the position $\Bx_0$ with strength $S_0$. 
%We then have that $U_{00}$ is given by 
%\begin{equation}
%U_{00}(\Bx) = \frac{S_0}{\ell}G_D(\Bx,\Bx_0) \ .)
%\end{equation}

The Green's function $G_D$ obeys
\begin{equation}
-\frac{1}{3}\ell^* \lap_\Bx G_D(\Bx,\Bx') + \mu_a G_D(\Bx,\Bx') =  \delta(\Bx-\Bx')
\end{equation}
along with the homogeneous boundary condition
\begin{equation}
G_D(\Bx,\Bx') + \ell \hat{\Bn}\cdot \grad_\Bx G_D(\Bx,\Bx') = 0 \ .
\end{equation}
It can be seen that the half-space Green's function, denoted $G_D^{(0)}$ can be expanded into plane waves of the form~\cite{markel_2004}
\begin{equation}
G_D^{(0)}(\Bx,\Bx') = \int \frac{d^2p}{(2\pi)^2} e^{i\Bp\cdot(\Brho-\Brho')} g(z,z';\Bp) \ ,
\end{equation}
where $\Bx =(\Brho,z)$ and
\begin{align}
&g(z,z';\Bp) = A_1(\Bp)e^{-\lambda({\Bp})|z-z'|}+A_2(\Bp) e^{-\lambda(\Bp)|z+z'|}
 \ , \\ 
&A_1(\Bp) = \frac{1}{2\lambda(\Bp)} \ , \quad A_2(\Bp) = \frac{1-\lambda(\Bp)\ell}{2\lambda(\Bp)(1+\lambda(\Bp)\ell)} \ , \\
&\lambda(\Bp) = \sqrt{\kappa^2+p^2} \ .
\end{align}

%Using the above result, we see that $U_{01}$ and $U_{11}$ can be obtained by replacing $G_D$ by $G_D^{(0)}$ in (\ref{int_U01}) and (\ref{int_U11}), respectively. Carrying out the required integrals we obtain
%\begin{equation}
%U_{01}(\Bx) = i c_{01}[F_{-1}(\Bx)+F_1(\Bx)]
%\end{equation}
%and
%\begin{equation}
%U_{11}(\Bx) =c_{11}[H_{1,1}(\Bx)+H_{1,-1}(\Bx)+H_{-1,1}(\Bx)+H_{-1,-1}(\Bx)] . 
%\end{equation}
%Here $F_m$ and $H_{mn}$ are defined in Appendix D, $c_{01} =(3\gamma \varepsilon k_0 S_0)/(8\ell^* \ell^2),$ and $c_{11} = (9 \gamma^2 \varepsilon_0^2 k_0^2S_0)/(32(\ell^*)^2\ell^2).$

\subsection{Point absorber}
We now consider the effect of a small absorbing inhomogeneity. The absorption coefficient is taken to be
\begin{equation}
\label{point_absorber}
\mu_a(\Bx) = \bar\mu_a + \delta\mu_a V\delta(\Bx-\Bx_1) \ ,
\end{equation}
where $\bar\mu_a$ is constant and $\delta\mu_a$, $V$, and $\Bx_1$ are the strength of the absorber, its volume and position, respectively.  The Green's function $G_D$ obeys the integral equation
\begin{align}
\label{int_eq}
G_D(\Bx,\Bx') = G_D^{(0)}(\Bx,\Bx') + \int d^3y G_D^{(0)}(\Bx,\By)\eta(\By)G_D(\By,\Bx') \ ,
\end{align}
where $\eta=\mu_a - \bar\mu_a$. If the absorber is relatively weak, so that $\delta\mu_a \ll \bar\mu_a$, then we can calculate the Green's function $G_D$ by making use of the Born approximation. We thus obtain
\begin{align}
\label{Born}
G_D(\Bx,\Bx') = G_D^{(0)}(\Bx,\Bx') + \delta\mu_a V G_D^{(0)}(\Bx,\Bx_1)G_D^{(0)}(\Bx_1,\Bx') \ ,
\end{align}
where we have replaced $G_D$ on the right hand side of (\ref{int_eq}) with $G_D^{(0)}$. Using this result, along with 
\begin{equation}
U_{00}(\Bx) = \frac{S_0}{\ell}G_D(\Bx,\Bx_0) \ ,
\end{equation}

\begin{figure}[t] 
\begin{center}
\includegraphics[width=0.4\textwidth]{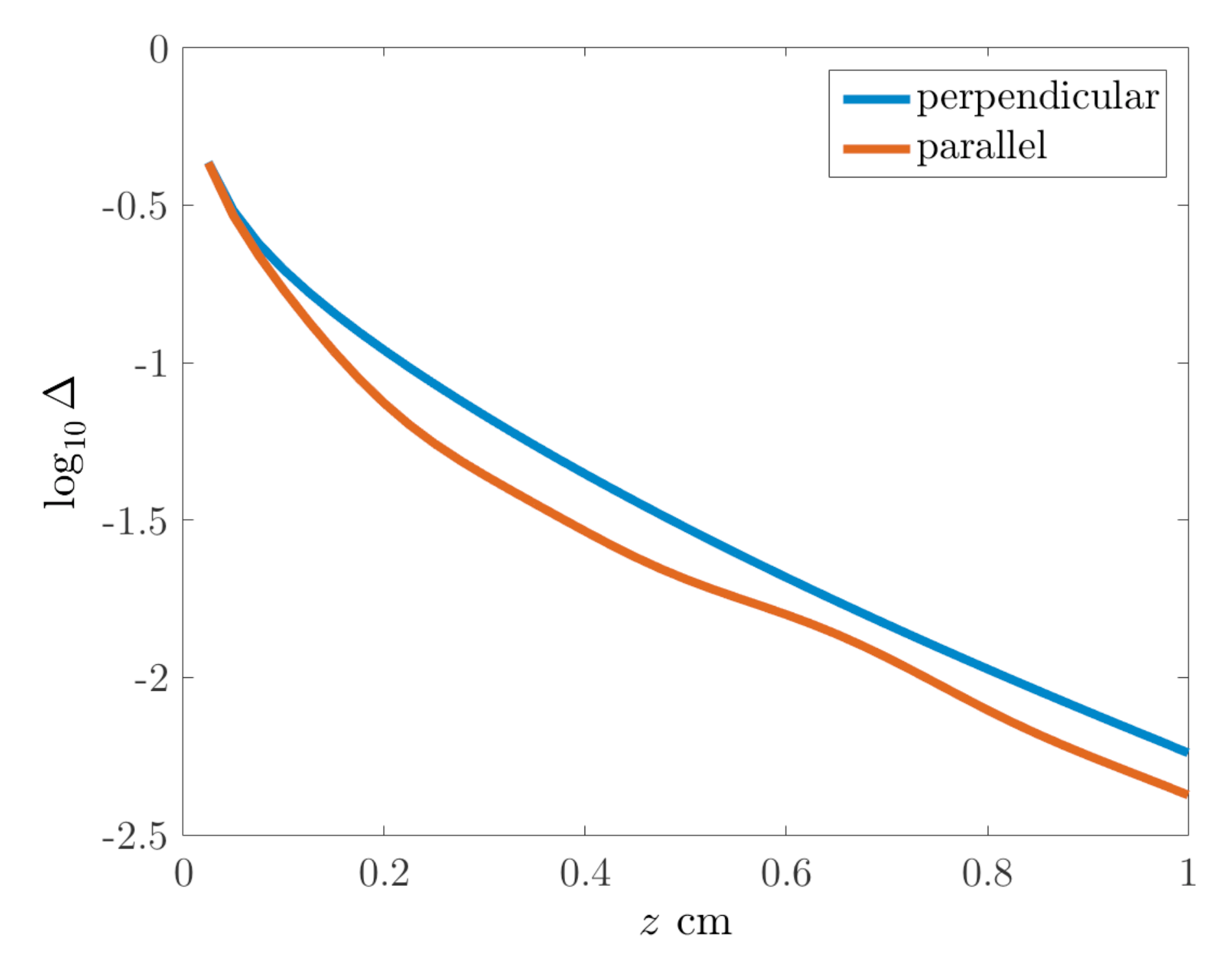} 
\end{center}
\caption{(Color online) $\Delta$ as a function of the depth $z_1$ of the absorber  for parallel and perpendicular orientations of the acoustic wavevector $\BQ$.}
\end{figure}
we can now calculate $U_{01}$ and $U_{11}$ from (\ref{int_U01}) and (\ref{int_U11}), respectively. For simplicity, we assume that the source and detector are located at the origin. We then find that
\begin{equation}
\label{U010}
\begin{split}
U_{01}(0) = ic_{01}\sum_{m}\left[F_m({0})-2\alpha\kappa^2\,{\rm Re}\, G_D({0},\Bx_1) F_m(\Bx_1) \right]
\end{split}
\end{equation}
and
\begin{eqnarray}
\label{U110}
\nonumber
U_{11}({0}) = c_{11} \sum_{m,n} \hspace{-10pt} &&\big[H_{mn}({0})-2 \alpha k_0^2 \,{\rm Re}\, \left\{G_D(0,\Bx_1)H_{mn}(\Bx_1)\right\} \\
&-&\alpha k_0^2 F_m(\Bx_1) F_n^*(\Bx_1)\big] ,
\end{eqnarray}
where $\alpha = (\delta \mu_a V)/\bar\mu_a$. Here $F_m$ and $H_{mn}$ are defined in Appendix D, $c_{01} =3\gamma \varepsilon k_0 S_0/8\ell^* \ell^2,$ and $c_{11} = 9 \gamma^2 \varepsilon_0^2 k_0^2S_0/32\ell^*{^2}\ell^2.$

We now estimate the sensitivity of the acousto-optic measurement to the presence of a small absorbing object. We work in the half-space geometry, in which the source and detector coincide, and are collinear with the point absorber. We define the relative change in intensity as
\begin{equation}
\Delta = \frac{\left|U_{11}-U_{11}(\delta\mu_a=0)\right|}{\left|U_{11}(\delta\mu_a=0)\right|} \ .
\end{equation}
The quantity $\Delta$ can be interpreted as the precision with which the intensity $I_{11}\propto U_{11}$ can be measured relative to the intensity in the absence of the absorber. For a fixed value of  $\Delta$, we can then estimate the threshold for the detection of the absorbing object, namely if $\Delta$ exceeds the experimental noise level we will say that an object is detectable.

Figure~3 shows a plot of $\Delta$ as a function of the distance $z_1$ of the absorber from the source and detector, for different values of the absorption contrast $\delta\mu_a/\bar\mu_a$. We consider separately the cases where the acoustic wave vector $\BQ$ is parallel or perpendicular to the line containing the source and detector. We see that for a noise level $\Delta=1\%$, it is possible to detect the object at a depth of 0.9 cm. We note that the parallel orientation of $\BQ$ is more favorable. At lower contrast, the depth at which the object can be detected decreases, while at higher contrast, the depth increases (not shown).

\section{Discussion}

We have derived the radiative transport equations that govern the acousto-optic effect in a random medium. Several comments on our results are necessary. First, the regime $\epsilon \ll \delta$, which corresponds to large-amplitude pressure waves, requires a theory that accounts for hydrodynamic interactions. Such interactions introduce short-range correlations in the susceptibility that would necessitate an analysis beyond the theory we have presented. Second, effects due to polarization of the optical field have not been discussed. Recent progress on polarized radiative transport and diffusion may lead to new results in this direction~\cite{borcea,carminati}.
Third, the detection thresholds we have obtained must be considered to be best-case estimates. We have not directly considered the effects of systematic errors in positioning of the source and detector or other experimental parameters. Fourth, when the medium is not known to consist of isolated inhomogeneities, it is of interest to recover the spatial dependence of the absorption. This inverse problem has so far only been studied for the case of the incoherent acousto-optic effect~\cite{Bal_2010,Ammari_2014_1,Ammari_2014_2,Ammari_2013,BalSchotland_PhysRev2014,BalMoskow,BCS,chung}. This paper provides the necessary radiative transport and diffusion equations to study the coherent problem.
These and other topics will be the subjects of future works.

\section*{Acknowledgments}

Valuable discussions with Guillaume Bal, Claude Boccara and Francis Chung are gratefully acknowledged. This work was supported in part by the NSF grant DMS-1619907.

\appendix

\begin{widetext}
\section*{Appendix A}
In this Appendix we derive (\ref{L1})--(\ref{L3}).
To proceed, we introduce the Wigner transform of the vectors $F$ and $G$ by
\begin{equation}
W[F,G]({\bf x},{\bf k}) = \int e^{-i{\bf y}\cdot{\bf k}} F\left({\bf x}-\frac{\epsilon}{2}{\bf y}\right) G^\dag\left({\bf x}+\frac{\epsilon}{2}{\bf y}\right)\,{d}^3 {y}.
\end{equation}
We also define $F^{(\pm)} = F({\bf x}\pm\frac{\epsilon}{2}{\bf y}).$
If $\phi_\epsilon$ is a solution of (\ref{phi_helmholtz}) we find that
\begin{equation}\label{eq:long_Wigner}
\begin{split}
0 = W[\epsilon^2\lap& \phi_\epsilon + k_0^2\left(\varepsilon_0 + 4\pi \sqrt\epsilon\eta(\Bx/\epsilon)\right) \phi_\epsilon +\frac{\epsilon k_0^2}{2}
\left(\gamma\varepsilon_0 + 4\pi \sqrt\epsilon\eta(\Bx/\epsilon)\right)\cos(\BQ\cdot\Bx)A\phi_\epsilon\,,\,\phi_\epsilon]-\\
&W[\phi_\epsilon, \,\epsilon^2\lap \phi_\epsilon + k_0^2\left(\varepsilon_0 + 4\pi \sqrt\epsilon\eta(\Bx/\epsilon)\right) \phi_\epsilon +\frac{\epsilon k_0^2}{2}
\left(\gamma\varepsilon_0 + 4\pi \sqrt\epsilon\eta(\Bx/\epsilon)\right)\cos(\BQ\cdot\Bx)A\phi_\epsilon].
\end{split}
\end{equation}
We now  write (\ref{eq:long_Wigner}) in terms of $W_\epsilon({\bf x},{\bf k}) = W[\phi_\epsilon,\phi_\epsilon]({\bf x},{\bf k}),$ where $\phi_\epsilon$ satisfies (\ref{phi_helmholtz}). First, we observe that
\begin{equation}
\left(\Delta \phi^{(-)}_\epsilon\right) {\phi^{(+)}_\epsilon}^\dag = \frac{1}{4} \left(\nabla_{\bf x}-\frac{2}{\epsilon}\nabla_{\bf y}\right)^2 \phi^{(-)}_\epsilon {\phi^{(+)}_\epsilon}^\dag,
\end{equation}
and
\begin{equation}
\phi_{\epsilon}^{(-)} \left(\Delta \phi^{(+)}_\epsilon\right)^\dag = \frac{1}{4} \left(\nabla_{\bf x}+\frac{2}{\epsilon}\nabla_{\bf y}\right)^2 \phi^{(-)}_\epsilon {\phi^{(+)}_\epsilon}^\dag .
\end{equation}
Thus
\begin{equation}
W[\Delta \phi_\epsilon,\phi_\epsilon]-W[\phi_\epsilon,\Delta \phi_\epsilon] = -2i\frac{\bf k}{\epsilon}\cdot\nabla_{\bf x}W_\epsilon.
\end{equation}
Next, we observe that 
\begin{equation}
W\left[\eta \phi_\epsilon, \phi_\epsilon\right]= \int \frac{e^{i{\bf p}\cdot{\bf x}/\epsilon}}{(2\pi)^3} \hat{\eta}({\bf p}) W_\epsilon\left({\bf x},{\bf k}+\frac{\bf p}{2}\right) {d}^3 {p},
\end{equation}
and
\begin{equation}
 W\left[\phi_\epsilon, \eta \phi_\epsilon\right]= \int \frac{e^{i{\bf p}\cdot{\bf x}/\epsilon}}{(2\pi)^3} \hat{\eta}({\bf p}) W_\epsilon\left({\bf x},{\bf k}-\frac{\bf p}{2}\right) {d}^3 {p}.
\end{equation}
In addition, for a constant matrix $A$ we have
\begin{equation}
\begin{split}
W\left[\cos({\bf Q}\cdot{\bf x})A \phi_\epsilon,\phi_\epsilon \right]({\bf x},{\bf k})&= \int {e^{i{\bf p}\cdot{\bf x}}}\frac{1}{2}\left[ \delta({\bf p}-{\bf Q})+\delta({\bf p}+{\bf Q})\right] A W_\epsilon({\bf x},{\bf k}+\frac{\epsilon}{2}{\bf p})\,{d}^3{p}\\
&=\frac{1}{2} A\,\left[ e^{i{\bf Q}\cdot {\bf x}}W_\epsilon\left({\bf x},{\bf k}+\frac{\epsilon}{2}{\bf Q}\right) + e^{-i{\bf Q}\cdot {\bf x}}W_\epsilon\left({\bf x},{\bf k}-\frac{\epsilon}{2}{\bf Q}\right) \right] \ .
\end{split}
\end{equation}
Likewise
\begin{equation}
W\left[\phi_\epsilon,\cos({\bf Q}\cdot{\bf x}) A\phi_\epsilon \right]({\bf x},{\bf k}) = \frac{1}{2} \left[e^{-i{\bf Q}\cdot {\bf x}}W_\epsilon\left({\bf x},{\bf k}+\frac{\epsilon}{2}{\bf Q}\right) + e^{i{\bf Q}\cdot {\bf x}}W_\epsilon\left({\bf x},{\bf k}-\frac{\epsilon}{2}{\bf Q}\right) \right]A^\dag.
\end{equation}
Finally we have,
\begin{equation}
\begin{split}
W\left[\eta\cos({\bf Q}\cdot{\bf x})A\phi_\epsilon, \phi_\epsilon\right] &= A\int\frac{e^{i{\bf p}\cdot{\bf x}}}{(2\pi)^3}\frac{e^{i{\bf p}'\cdot{\bf x}/\epsilon}}{(2\pi)^3} \hat{\eta}({\bf p}') \frac{(2\pi)^3}{2}\left[ \delta({\bf p}-{\bf Q})+\delta({\bf p}+{\bf Q})\right]\,W_\epsilon\left({\bf x},{\bf k}+\frac{{\bf p}'}{2}+\frac{\epsilon}{2}{\bf p}\right) {d}^3{ p}\,{d}^3{p}',\\
&=\frac{1}{2} A\int \frac{e^{i{\bf p}'\cdot{\bf x}/\epsilon}}{(2\pi)^3} \hat{\eta}({\bf p}')\left[e^{i{\bf Q}\cdot {\bf x}}W_\epsilon\left({\bf x},{\bf k}+\frac{\bf p'}{2}+\frac{\epsilon}{2}{\bf Q}\right)+e^{-i{\bf Q}\cdot {\bf x}}W_\epsilon\left({\bf x},{\bf k}+\frac{\bf p'}{2}-\frac{\epsilon}{2}{\bf Q}\right) \right] {d}^3 {p}',
\end{split}
\end{equation}
and
\begin{equation}
\begin{split}
W\left[\phi_\epsilon, \eta\cos({\bf Q}\cdot{\bf x})A\phi_\epsilon\right] &=\frac{1}{2} \int \frac{e^{i{\bf p}'\cdot{\bf x}/\epsilon}}{(2\pi)^3} \hat{\eta}({\bf p}')\left[e^{i{\bf Q}\cdot {\bf x}}W_\epsilon\left({\bf x},{\bf k}-\frac{\bf p'}{2}+\frac{\epsilon}{2}{\bf Q}\right)+e^{-i{\bf Q}\cdot {\bf x}}W_\epsilon\left({\bf x},{\bf k}-\frac{\bf p'}{2}-\frac{\epsilon}{2}{\bf Q}\right) \right]A^\dag {d}^3{p}'.
\end{split}
\end{equation}
Applying the above identities to (\ref{eq:long_Wigner}), we see that $W_\epsilon$ satisfies
\begin{equation}\label{eq:Liouville}
{\bf k}\cdot \nabla W_\epsilon({\bf x},{\bf k}) +\left( \frac{1}{\sqrt{\epsilon}}L_1+L_2+\sqrt{\epsilon}L_3\right)W_\epsilon({\bf x},{\bf k})=0,
\end{equation}
where
\begin{equation}\label{eq:Liouville_def}
\begin{split}
L_1Z&=2\pi i k_0^2 \int \frac{e^{i{\bf p}\cdot{\bf x}/\epsilon}}{(2\pi)^3} \hat{\eta}({\bf p}) \left[Z\left({\bf x},{\bf k}+\frac{\bf p}{2}\right)-Z\left({\bf x},{\bf k}-\frac{\bf p}{2}\right)\right] {d}^3 {p},\\
L_2Z&=\frac{\gamma \varepsilon_0 i}{8}k_0^2A\left[ e^{i{\bf Q}\cdot {\bf x}}Z\left({\bf x},{\bf k}+\frac{\epsilon}{2}{\bf Q}\right) + e^{-i{\bf Q}\cdot {\bf x}}Z\left({\bf x},{\bf k}-\frac{\epsilon}{2}{\bf Q}\right) \right]\\
&\,\, -\frac{\gamma \varepsilon_0 i}{8}k_0^2\left[ e^{-i{\bf Q}\cdot {\bf x}}Z\left({\bf x},{\bf k}+\frac{\epsilon}{2}{\bf Q}\right) + e^{i{\bf Q}\cdot {\bf x}}Z\left({\bf x},{\bf k}-\frac{\epsilon}{2}{\bf Q}\right) \right]A^\dag,\\
L_3Z&=\frac{ \pi i}{2}k_0^2 A \int \frac{e^{i{\bf p}\cdot{\bf x}/\epsilon}}{(2\pi)^3} \hat{\eta}({\bf p})\left[e^{i{\bf Q}\cdot {\bf x}}Z\left({\bf x},{\bf k}+\frac{\bf p}{2}+\frac{\epsilon}{2}{\bf Q}\right)+e^{-i{\bf Q}\cdot {\bf x}}Z\left({\bf x},{\bf k}+\frac{\bf p}{2}-\frac{\epsilon}{2}{\bf Q}\right) \right] {d}^3 {p}\\
&\,\,-\frac{ \pi i}{2}k_0^2 \int \frac{e^{i{\bf p}\cdot{\bf x}/\epsilon}}{(2\pi)^3} \hat{\eta}({\bf p})\left[e^{-i{\bf Q}\cdot {\bf x}}Z\left({\bf x},{\bf k}+\frac{\bf p}{2}+\frac{\epsilon}{2}{\bf Q}\right)+e^{i{\bf Q}\cdot {\bf x}}Z\left({\bf x},{\bf k}+\frac{\bf p}{2}-\frac{\epsilon}{2}{\bf Q}\right) \right] A^\dag {d}^3 {p}.
\end{split}
\end{equation}

\section*{Appendix B}
In this Appendix we derive (\ref{pre_RTE}). We begin by defining
\begin{equation}\label{eq:W1_solved}
\tilde{W}_1(\Bx,\Bq,\Bk) = 2\pi k_0^2 \tilde{\eta}(\Bq) \frac{W_0(\Bx,\Bk+\Bq/2)-W_0(\Bx,\Bk-\Bq/2)}{\Bq\cdot\Bk+i\theta},
\end{equation}
where $\theta$ is a small positive regularization parameter. Note that if
\begin{equation}
\langle \eta(\Br) \eta(\Br') \rangle = C(|\Br-\Br'|),
\end{equation}
then
\begin{equation}\label{eq:FT_correlations}
\langle \tilde{\eta}(\Bk) \tilde{\eta}(\Bk') \rangle = 2 \pi^3 \delta(\Bk+\Bk')\tilde{C}(|\Bk-\Bk'|/2).
\end{equation}
We define the quantity $T_\pm$  by
\begin{equation}
T_\pm = \int \frac{d^3p}{(2\pi)^3} e^{i\Bp\cdot\Bx/\epsilon} \langle\tilde\eta(\Bp)W_1(\Bx,\Bq,\Bk\pm\Bp/2)\rangle.
\end{equation}
From (\ref{eq:W1_solved}) it follows that
\begin{equation}\label{eq:Tpm_substituted}
T_\pm =2\pi k_0^2 \int \frac{d^3p\, d^3q}{(2\pi)^6} e^{i(\Bp+\Bq)\cdot{\bf \xi}}\langle\tilde\eta(\Bp) \eta(\Bq) \rangle\frac{W_0(\Bx,\Bk\pm \Bp/2+\Bq/2)-W_0(\Bx,\Bk\pm \Bp/2-\Bq/2)}{\Bq\cdot(\Bk\pm\Bp/2)+i\theta}.
\end{equation}
Upon substitution of (\ref{eq:FT_correlations}) into (\ref{eq:Tpm_substituted}) we obtain
\begin{equation}
T_\pm = \pm 2\pi k_0^2 \int \frac{d^3p}{(2\pi)^3} \tilde{C}(|\Bp|) \frac{W_0(\Bx,\Bk)-W_0(\Bx,\Bk+\Bp)}{\mp\Bp\cdot(\Bk+ \Bp/2)+i\theta}.
\end{equation}
Making the transformation $\Bp \rightarrow \Bp-\Bk$ we find
\begin{equation}
T_\pm = \pm 2\pi k_0^2 \int \frac{d^3p}{(2\pi)^3} \tilde{C}(|\Bk - \Bp|) \frac{W_0(\Bx,\Bk)-W_0(\Bx,\Bp)}{\pm (k^2-p^2)/2+i\theta},
\end{equation}
from which we see that
\begin{equation}
T_+-T_- = 2\pi k_0^2  \int \frac{d^3p}{(2\pi)^3} \tilde{C}(|\Bk - \Bp|) \left[W_0(\Bx,\Bk)-W_0(\Bx,\Bp)\right] \frac{-(2\theta)i}{(\theta)^2+(k^2-p^2)^2/4} \ .
\end{equation}
Using the fact that
\begin{equation}
\lim_{\theta \rightarrow 0} \frac{\theta}{\theta^2+x^2} = \pi \delta(x)
\end{equation}
we obtain
\begin{equation}
T_+-T_- = 4 \pi^2 i k_0^2  \int \frac{d^3p}{(2\pi)^3}\delta\left(\frac{k^2}{2}-\frac{p^2}{2}\right) \tilde{C}(|\Bk - \Bp|) \left[W_0(\Bx,\Bp)-W_0(\Bx,\Bk)\right] ,
\end{equation}
which corresponds to the right hand side of (\ref{pre_RTE}).

\section*{Appendix C}
In this Appendix we derive (\ref{U11_integral}).
We proceed by taking the Fourier transform of (\ref{int_U01}), noting that
\begin{equation}
\tilde{G}_D({\Bk}) = \frac{3}{\ell^*}\frac{1}{k^2 + \kappa^2} \ .
\end{equation}
Thus
\begin{equation}\label{eq:U_01_FT}
\begin{split}
\tilde{U}_{01}(\Bk) =  \frac{i\gamma\varepsilon_0k_0}{8}e^{-i\Bk\cdot\Bx_0}\sum_{m} &e^{im\BQ\cdot\Bx_0} \frac{1}{k^2+\kappa^2}
 \frac{1}{(\Bk-m\BQ)^2+\kappa^2} ,
\end{split}
\end{equation}
where the sum is over $m\in\{-1,1\}$.
We next observe that for positive constants $A$ and $B,$
\begin{equation}
\frac{1}{AB} = \int_0^1 {d}s  \frac{1}{[As+(1-s)B]^2}.
\end{equation}
Upon application of this identity and taking an inverse Fourier transform, we find that (\ref{eq:U_01_FT}) becomes
\begin{equation}
\begin{split}
{U}_{01}(\Bx) =&  \frac{i\gamma\varepsilon_0k_0}{8(2\pi)^3}\sum_{m}e^{im\BQ\cdot\Bx_0}\int_0^1 {d}s \int {d}^3 k \,\, e^{i\Bk\cdot(\Bx-\Bx_0)} \frac{1}{[(k^2+\kappa^2)(1-s)+s(\Bk-m\BQ)^2+s\kappa^2]^2}.
\end{split}
\end{equation}
Next, let $\Br = \Bx-\Bx_0,$ ${\Bk}' = \Bk-s m \BQ$ and 
\begin{equation}
\beta(s) = \sqrt{\kappa^2+m^2\BQ^2 s(1-s^2)}.
\end{equation}
Then
\begin{equation}
\begin{split}
{U}_{01}(\Bx) =& \frac{i\gamma\varepsilon_0k_0}{8(2\pi)^3}\sum_{m} e^{im\BQ\cdot\Bx_0}
\int_0^1 {d}s\, e^{i\Br\cdot sm\BQ}\int {d}^3 k' \,\, e^{i\Bk'\cdot\Br} \frac{1}{({\Bk'}^2+\beta^2)^2}.
\end{split}
\end{equation}
Performing the integral over $\Bk'$ we obtain
\begin{equation}
\begin{split}
{U}_{01}(\Bx) &= \frac{i\gamma\varepsilon_0k_0}{64\pi}\sum_{m} e^{im\BQ\cdot\Bx_0}\int_0^1 {d}s\, e^{ism\BQ\cdot\Br} \frac{e^{-\beta(s)r}}{\beta(s)}.
\end{split}
\end{equation}
Similarly, we can obtain an expression for $U_{11}$ of the form
\begin{equation}\label{eq:U_11_FT}
\begin{split}
U_{11}(\Bx) &= \frac{\gamma \varepsilon_0 k_0}{2}\int {d}^3 x' G_D(\Bx-\Bx') \cos(\BQ\cdot\Bx') {\rm Im}U_{01}(\Bx')\\
 &=\frac{(\gamma \varepsilon_0 k_0)^2}{8}\int\int {d}^3 x'\,{d}^3 x'' G_D(\Bx-\Bx') \cos(\BQ\cdot\Bx')G_D(\Bx'-\Bx'')\cos(\BQ\cdot\Bx'')G_D(\Bx''-\Bx_0) \\
 &=\frac{(\gamma \varepsilon_0 k_0)^2}{32}\sum_{m,n}\int {d}^3 x'\,{d}^3 x'' G_D(\Bx-\Bx') e^{im\BQ\cdot\Bx'}G_D(\Bx'-\Bx'')e^{in\BQ\cdot\Bx''}G_D(\Bx''-\Bx_0)\\
 &=\frac{(\gamma \varepsilon_0 k_0)^2}{32} \sum_{m,n}e^{i\BQ\cdot(m\Bx+n\Bx_0)}\int \frac{{d}^3 k}{(2\pi)^3}\frac{1}{(\Bk+m\BQ)^2+\kappa^2}\frac{1}{k^2+\kappa^2}\frac{1}{(\Bk-n\BQ)^2+\kappa^2}e^{i\Bk\cdot(\Bx-\Bx_0)}.
\end{split}
\end{equation}
Next, we observe that for positive constants $A,B$ and $C$,
\begin{equation}\label{eq:Feynman_3D}
\frac{1}{ABC} = 2 \int_0^1{d}x \, \int_0^{1-x}{d}y \frac{1}{[Ax+By+C(1-x-y)]^3}.
\end{equation}
Upon substitution of (\ref{eq:Feynman_3D}) into the final line of (\ref{eq:U_11_FT}) and letting $\Br = \Bx-\Bx_0,$ we obtain
\begin{equation}\label{eq:U_11_after_feyn}
U_{11}(\Bx)=\frac{(\gamma \varepsilon_0 k_0)^2}{16}\sum_{m,n} e^{i\BQ\cdot(m\Bx+n\Bx_0)} \int_0^1 {d}s\, \int_0^{1-s}{d}t\, \int \frac{{d}^3 k}{(2\pi)^3} e^{i\Bk\cdot\Br}\frac{1}{[k^2+2\Bk\cdot\Bp +p^2+\gamma^2(s,t)]^3},
\end{equation}
where $\Bp = (ms-nt)\BQ$ and 
\begin{equation}
\gamma(s,t) =\sqrt{ \kappa^2+ (s m^2+t n^2) Q^2 -{\bf p}^2(s,t)} = \sqrt{\kappa^2+(s+t)Q^2-{\bf p}^2(s,t)} .
\end{equation}
Changing variables in (\ref{eq:U_11_after_feyn}) yields
\begin{equation}
\begin{split}
U_{11}(\Bx)&=\frac{(\gamma \varepsilon_0 k_0)^2}{16}\sum_{m,n} e^{i\BQ\cdot(m\Bx+n\Bx_0)} \int_0^1 {d}s\, \int_0^{1-s}{d}t\, e^{-i\Bp(s,t) \cdot \Br}\int \frac{{d}^3 k}{(2\pi)^3} e^{i\Bk\cdot\Br}\frac{1}{(k^2+\gamma^2)^3}\\
&=\frac{(\gamma \varepsilon_0 k_0)^2}{512 \pi}\sum_{m,n} e^{i\BQ\cdot(m\Bx+n\Bx_0)} \int_0^1 {d}s\, \int_0^{1-s}{d}t\, e^{-i\Bp(s,t) \cdot \Br}	\frac{e^{-\gamma r}}{\gamma^2}\left(r+\frac{2}{\gamma}\right).\\
\end{split}
\end{equation}

\section*{Appendix D}
In this Appendix we derive (\ref{U010}) and (\ref{U110}). Substituting (\ref{point_absorber}) and (\ref{Born}) into (\ref{int_U01}) and (\ref{int_U11}), respectively we obtain 
\begin{equation}
\begin{split}
&U_{00}(\Bx) = \frac{S_0}{\ell^2}\left[G_D(\Bx,{0})-\kappa^2\alpha G_D(\Bx,\Bx_1)G_D(\Bx_1,0)\right],\\
&U_{01}(\Bx) = \frac{3\gamma \varepsilon_0 k_0S_0}{4 \ell^* \ell^2}\left[\int_{z'\ge0}d^3x' G_D(\Bx,\Bx')\cos(\BQ\cdot \Bx')G_D(\Bx',{0})- \alpha\kappa^2G_D(\Bx,\Bx_1) \int_{z'\ge0}d^3x' G_D(\Bx_1,\Bx')\cos(\BQ\cdot \Bx')G_D(\Bx',{0})\right]\\
&\hspace{3 cm}-\alpha\kappa^2\frac{3\gamma \varepsilon_0 k_0S_0}{4 \ell^* \ell^2}\int_{z'\ge0}d^3x' G_D(\Bx,\Bx')\cos(\BQ\cdot \Bx')G_D(\Bx',\Bx_1)G_D(\Bx_1,0) ,\\
&U_{11}(\Bx) = \frac{9 \gamma^2 \varepsilon_0^2 k_0^2S_0}{8 (\ell^*)^2 \ell^2}\left[\int_{z'\ge0}d^3x' d^3x'' G_D(\Bx,\Bx')\cos(\BQ\cdot \Bx')G_D(\Bx',\Bx'')\cos(\BQ\cdot\Bx'')G_D(\Bx'',{0}) \right]\\
&\hspace{3 cm}-\alpha\kappa^2\frac{9 \gamma^2 \varepsilon_0^2 k_0^2S_0}{8 (\ell^*)^2 \ell^2}G_D(\Bx,\Bx_1)\int_{z'\ge0}d^3x' \,\int_{z''\ge0}d^3x'' G_D(\Bx_1,\Bx')\cos(\BQ\cdot \Bx')G_D(\Bx',\Bx'')\cos(\BQ\cdot\Bx'')G_D(\Bx'',{0})\\
&\hspace{3 cm}-\alpha\kappa^2\frac{9 \gamma^2 \varepsilon_0^2 k_0^2S_0}{8 (\ell^*)^2 \ell^2}\int_{z'\ge0}d^3x' \,\int_{z''\ge0}d^3x'' G_D(\Bx,\Bx')\cos(\BQ\cdot \Bx')G_D(\Bx',\Bx'')\cos(\BQ\cdot\Bx'')G_D(\Bx'',\Bx_1)G_D(\Bx_1,{0})\\
&\hspace{3 cm}-\alpha\kappa^2\frac{9 \gamma^2 \varepsilon_0^2 k_0^2S_0}{8 (\ell^*)^2 \ell^2}\int_{z'\ge0}d^3x' \,G_D(\Bx,\Bx')\cos(\BQ\cdot \Bx')G_D(\Bx',\Bx_1)\int_{z''\ge0}d^3x'' G_D(\Bx_1,\Bx'')\cos(\BQ\cdot\Bx'')G_D(\Bx'',{0})
 .
\end{split}
\end{equation}
To proceed further we write $\BQ = (\BQ_\perp,Q_\parallel),$ and define the functions $F_m$ and $H_{mn}$ by
\begin{equation}
\begin{split}
&F_m(\Bx) = \int_{z'\ge0}d^3x' G_D(\Bx,\Bx') e^{im Q_\parallel z'}e^{im\BQ_\perp \cdot \Brho'} G_D(\Bx',{0}),\\
&H_{mn}(\Bx) = \int_{z'\ge0}d^3x'\,\int_{z''\ge0}d^3x'' \,G_D(\Bx,\Bx') e^{im Q_\parallel z'}e^{im\BQ_\perp \cdot \Brho'} G_D(\Bx',\Bx'')e^{in Q_\parallel z''}e^{in\BQ_\perp \cdot \Brho''} G_D(\Bx'',{0}).\\
\end{split}
\end{equation}
Observe that $G_D(\Bx,\Bx') = G_D^*(\Bx',\Bx)$
and thus
\begin{equation}
\begin{split}
 \int_{z'\ge0}d^3x' G_D(0,\Bx') e^{im Q_\parallel z'}e^{im\BQ_\perp \cdot \Brho'} G_D(\Bx',\Bx)&= \int_{z'\ge0}d^3x' G_D^*(\Bx,\Bx') e^{im Q_\parallel z'}e^{im\BQ_\perp \cdot \Brho'} G_D^*(\Bx',0)\\
 &= F_{-m}^*(\Bx).
 \end{split}
\end{equation}
Similarly,
\begin{equation}
\begin{split}
\int_{z'\ge0}d^3x'\,\int_{z''\ge0}d^3x'' \,G_D(0,\Bx') e^{im Q_\parallel z'}e^{im\BQ_\perp \cdot \Brho'} G_D(\Bx',\Bx'')e^{in Q_\parallel z''}e^{in\BQ_\perp \cdot \Brho''} G_D(\Bx'',\Bx) = H^*_{-n,-m}(\Bx).
\end{split}
\end{equation}
Using the above results, we can write the data $U_{11}({0})$ as
\begin{equation}
U_{11}({0}) = U_{11}^{(0)}({0}) - \frac{\alpha \kappa^2 c}{4} \sum_{m,n}\left[G_D(\Bx_1,{0}) H_{mn}^*(\Bx_1)+G_D^*(\Bx_1,{0}) H_{mn}(\Bx_1) +F_m(\Bx_1)F_n^*(\Bx_1)\right], 
\end{equation}
where
$
c={9 \gamma^2 \varepsilon_0^2 k_0^2S_0/8 (\ell^*)^2 \ell^2}
$
and $U_{11}^{(0)} = \frac{c}{4}\sum_{m,n} H_{mn}(\Bx_1,{0})$ is the part of the measurement due solely to the homogeneous medium.

To evaluate $F_m$ we begin by observing that if $\lambda(\Bp) \neq \beta,$
\begin{equation}
\int_0^\infty {d}z' e^{-\lambda(p)|z-z'|} e^{-\beta z'} = \frac{2\lambda(p)e^{-\beta z}}{\lambda^2(p)-\beta^2 } - \frac{e^{-\lambda(p) z}}{\lambda(p)-\beta},
\end{equation}
and
\begin{equation}
\int_0^\infty {d}z' e^{-\lambda(p)|z+z'|} e^{-\beta z'} = \frac{e^{-\lambda(p)z}}{\lambda(p)+\beta}.
\end{equation}
Thus, if $\lambda(\Bp) \neq \beta,$
\begin{equation}
\int_0^\infty {d}z' g(z,z';\Bp) e^{-\beta z'} = L_\Bp(\beta) e^{-\beta z}+R_\Bp(\beta) e^{-\lambda(\Bp) z},
\end{equation}
where
\begin{equation}
L_\Bp^\beta = \frac{2\lambda(\Bp) A_1(\Bp)}{\lambda^2(\Bp)-\beta^2},  \quad R_\Bp^\beta = \frac{A_2(\Bp)}{\lambda(\Bp)+\beta} -\frac{A_1(\Bp)}{\lambda(\Bp)-\beta}.
\end{equation}
Additionally, if $\lambda(\Bp) = \beta,$ then
\begin{equation}
\int_0^\infty {d}z' g(z,z';\Bp) e^{-\beta z'} = [A_1(\Bp)+A_2(\Bp)]\frac{e^{-\lambda(\Bp)z}}{2\lambda(\Bp)}+A_1(\Bp) ze^{-\lambda(\Bp)z}.
\end{equation}
Hence, if $\Omega = \{ \Bp \in \mathbb{R}^2\,: \,  \Bp \neq \Bp \pm \BQ_\perp, \,\, \Bp \neq \Bp\pm2\BQ_\perp \},$
\begin{equation}
\begin{split}
F_m(\Bx) &= \int_0^\infty {d}z' \int d^2\rho'\int \frac{d^2p}{(2\pi)^2} \int \frac{d^3p'}{(2\pi)^2} e^{i\Bp\cdot(\Brho-\Brho')} e^{i\Bp'\cdot\Brho'} e^{im \BQ_\perp\cdot {\Brho'}} e^{imQ_\parallel z'} g(z,z';\Bp)\,g(z',0;\Bp')\\
&= \int_0^\infty {d}z'\int \frac{d^3p'}{(2\pi)^2} e^{i(\Bp'+m\BQ_\perp)\cdot\Brho}  e^{-(\lambda(\Bp')-imQ_\parallel) z'} g(z,z';\Bp'+m\BQ_\perp)\,[A_1(\Bp')+A_2(\Bp')]\\
&=\int_{\mathbb{R}^2\setminus \Omega} \frac{d^3p}{(2\pi)^2} e^{i(\Bp+m\BQ_\perp)\cdot\Brho}\,[A_1(\Bp)+A_2(\Bp)] 
\left[L_{\Bp+m\BQ_\perp}^{\lambda(\Bp)-imQ_\parallel} e^{-(\lambda(\Bp)-imQ_\parallel)z} +R_{\Bp+m\BQ_\perp}^{\lambda(\Bp)-imQ_\parallel}e^{-\lambda(\Bp+m\BQ_\perp)z}\right].\\
\end{split}
\end{equation}
Similarly,
\begin{equation}
\begin{split}
H_{mn}(\Bx) &= \int_0^\infty {d}z' \int d^2\rho'\int \frac{d^3p}{(2\pi)^2}e^{i\Bp\cdot(\Brho-\Brho')} e^{im \BQ_\perp\cdot \Brho'} e^{imQ_\parallel z'} g(z,z';\Bp)\,F_n(\Brho',z')\\
&=\int_0^\infty {d}z'\int_{\mathbb{R}^2 \setminus \Omega} \frac{d^3p}{(2\pi)^2} e^{i(\Bp+(m+n)\BQ_\perp)\cdot\Brho}\,[A_1(\Bp)+A_2(\Bp)]g(z,z';\Bp+(m+n)\BQ_\perp)\,e^{imQ_\parallel z'} \\ \times
&\left[L_{\Bp+n\BQ_\perp}^{\lambda(\Bp)-inQ_\parallel} e^{-(\lambda(\Bp)-inQ_\parallel)z'} +R_{\Bp+n\BQ_\perp}^{\lambda(\Bp)-inQ_\parallel}e^{-\lambda(\Bp+n\BQ_\perp)z'}\right].
\end{split}
\end{equation}
Note that if $m+n \neq 0,$
\begin{equation}
\begin{split}
H_{mn}(\Bx) &= \int_{\mathbb{R}^2 \setminus \Omega}\frac{d^3p}{(2\pi)^2} e^{i(\Bp+(m+n)\BQ_\perp)\cdot\Brho}\,[A_1(\Bp)+A_2(\Bp)] \\ \times
&\,\,\left\{L_{\Bp+(m+n)\BQ_\perp}^{\lambda(\Bp)-i(m+n)Q_\parallel}L_{\Bp+n\BQ_\perp}^{\lambda(\Bp)-inQ_\parallel} e^{-(\lambda(\Bp)-i(n+m)Q_\parallel)z}+R_{\Bp+(m+n)\BQ_\perp}^{\lambda(\Bp)-i(m+n)Q_\parallel}L_{\Bp+n\BQ_\perp}^{\lambda(\Bp)-inQ_\parallel} e^{-\lambda(\Bp+(m+n)\BQ_\perp)z} \right.\\
&\,\, + \left. L_{\Bp+(m+n)\BQ_\perp}^{\lambda(\Bp+n\BQ_\perp)-imQ_\parallel}R_{\Bp+n\BQ_\perp}^{\lambda(\Bp)-inQ_\parallel} e^{-(\lambda(\Bp+n\BQ_\perp)-imQ_\parallel)z}+R_{\Bp+(m+n)\BQ_\perp}^{\lambda(\Bp+n\BQ_\perp)-imQ_\parallel}R_{\Bp+n\BQ_\perp}^{\lambda(\Bp)-inQ_\parallel} e^{-\lambda(\Bp+(m+n)\BQ_\perp)z}\right\},
\end{split}
\end{equation}
and if $m+n = 0,$
\begin{equation}
\begin{split}
H_{mn}(\Bx) &= \int_{\mathbb{R}^2 \setminus \Omega}\frac{d^3p}{(2\pi)^2} e^{i\Bp\cdot\Brho}\,[A_1(\Bp)+A_2(\Bp)]\\ \times
&\,\,\left\{\frac{[A_1(\Bp)+A_2(\Bp)]}{2\lambda(\Bp)} L_{\Bp+n\BQ_\perp}^{\lambda(\Bp)-inQ_\parallel} e^{-(\lambda(\Bp)-i(n+m)Q_\parallel)z}+A_1(\Bp) L_{\Bp+n\BQ_\perp}^{\lambda(\Bp)-inQ_\parallel} e^{-\lambda(\Bp)z}z \right.\\
&\,\, \left. + L_{\Bp}^{\lambda(\Bp+n\BQ_\perp)-imQ_\parallel}R_{\Bp+n\BQ_\perp}^{\lambda(\Bp)-inQ_\parallel} e^{-(\lambda(\Bp+n\BQ_\perp)-imQ_\parallel)z}+R_{\Bp}^{\lambda(\Bp+n\BQ_\perp)-imQ_\parallel}R_{\Bp+n\BQ_\perp}^{\lambda(\Bp)-inQ_\parallel} e^{-\lambda(\Bp)z}\right\}.
\end{split}
\end{equation}

\end{widetext}

\end{document}